\documentclass[aps,prf,superscriptaddress,longbibliography]{revtex4-1}
\usepackage[latin1]{inputenc}
\usepackage{graphicx}
\usepackage{amsmath}
\usepackage{amssymb}
\usepackage{pifont}
\usepackage{epstopdf}

\def\b

\def\HS{\mbox{\scriptsize HS}}

\renewcommand{\Re}{\mathrm{Re}}

 \usepackage{color}

\newcommand{\eq}{\mathrm{eq}}
\newcommand{\zeq}{z^\mathrm{eq}}

\begin{document}

\title{Inertial migration of neutrally-buoyant particles in superhydrophobic channels}

 \author{Tatiana V. Nizkaya}
  \affiliation{A.N. Frumkin Institute of Physical Chemistry and
Electrochemistry, Russian Academy of Science, 31 Leninsky Prospect,
119071 Moscow, Russia}

 \author{Evgeny S. Asmolov}
\affiliation{A.N. Frumkin Institute of Physical Chemistry and
Electrochemistry, Russian Academy of Science, 31 Leninsky Prospect,
119071 Moscow, Russia}
\affiliation{M.V. Lomonosov Moscow State
University, 119991 Moscow, Russia}
\author{Jens Harting}
 \affiliation{Helmholtz Institute Erlangen-N\"urnberg for Renewable Energy, Forschungszentrum J\"ulich,\\ F\"urther Str. 248, 90429 N\"{u}rnberg, Germany}
 \affiliation{Department of Applied Physics, Eindhoven University of Technology, PO box 513, 5600MB Eindhoven, The Netherlands}
 \author{Olga I. Vinogradova}
\email[Corresponding author: ]{oivinograd@yahoo.com}
 \affiliation{A.N. Frumkin Institute of Physical Chemistry and
   Electrochemistry, Russian Academy of Science, 31 Leninsky Prospect,
   119071 Moscow, Russia}
 \affiliation{M.V. Lomonosov Moscow State
   University, 119991 Moscow, Russia}
 \affiliation{DWI - Leibniz Institute for Interactive Materials, Forckenbeckstr. 50, 52056 Aachen,
   Germany}

\date{\today}
\begin{abstract}
At finite Reynolds numbers particles migrate across flow streamlines to their equilibrium
positions in microchannels. Such a migration is attributed to an inertial lift force, and it is well-known that the equilibrium location of neutrally-buoyant particles is determined only by their size and the Reynolds number. Here we demonstrate that the decoration of a bottom wall of the channel by superhydrophobic grooves provides additional possibilities for manipulation of neutrally-buoyant particles. It is shown that the effective anisotropic hydrodynamic slip of such a bottom wall can be readily used to alter the equilibrium positions of particles and to generate their motion transverse to the pressure gradient. These results may guide the design of novel inertial microfluidic devices for efficient sorting of neutrally-buoyant microparticles by their size.

\end{abstract}
\maketitle
\section{Introduction}
\label{sec:introduction}

Inertial microfluidic systems  have been shown to be very useful
for focusing and separation of microparticles of different size, density and
shape with increased control and sensitivity, which is important for a wide
range of applications in chemistry, biology, and medicine~\citep{dicarlo07,bhagat08,gossett2010}.  Compared to classical microfluidic devices,
which usually rely on relatively small flow velocities (low-Reynolds-number
laminar flows), inertial microfluidics operates with higher rates when
fluid inertia becomes significant. At finite Reynolds numbers particles
experience an inertial lift force, which induces their migration across streamlines
of carrier flow to some equilibrium positions in microchannels. Thanks to this force, the focusing of particles in inertial microfluidic  devices could be achieved without applying any transverse external fields, that offers unique advantages in separation technologies. Since unlike field-flow fractionation methods, the inertial separation is based on focusing of particles at distinct equilibrium positions, but not on a difference in retention times~\cite{Giddings1976,Giddings1977} or in migration velocities~\cite{zhang1994separation}, it can be used for a continuous fractionation of particle dispersions at high
throughput (see
~\cite{zhang2016,stoecklein2018nonlinear} for recent reviews). An especially challenging problem is, naturally, the fractionation of spherical neutrally-buoyant microparticles, since it can only exploit the difference in their size. Therefore, the possibilities of controlled manipulations of such particles by using an inertial lift force remain quite limited.

Inertial separation of neutrally-buoyant particles implies that they should
focus at different locations, depending on their sizes. However, the dependence of the equilibrium positions on particle radii is too weak to be used in microfluidic applications. An efficient strategy for a separation of such particles on microscale is normally to balance
the inertial lift force by another one, which scales differently with the particle size. This includes such external forces as electric~\cite{zhang2014real} or magnetic~\cite{dutz2017fractionation}. The inertial lift can also be balanced by the Dean force due to a secondary rotational flow caused by inertia of the fluid itself, which can be generated in curved channels~\citep{dicarlo07,bhagat08}, leading to a migration of particles in vertical directions and altering their equilibrium positions. A promising direction is also to induce an additional force by exploiting porous channel walls. With such walls the drag to permeate flow could be used to control focusing positions of particles~\cite{altena1984,asmolov2011,garcia2017}.

Microfabrication has opened the possibility to elaborate channels whose surfaces are patterned in  a  very  well  controlled  way, thus  providing  properties that they did not have when flat or slightly disordered, and the best known example of such surfaces is probably superhydrophobic (SH) ones~\cite{quere.d:2008}. The large effective slip length (the distance within the solid at which the flow profile extrapolates to zero) of SH surfaces compared to simple, smooth channels can greatly lower the viscous drag and reduce the tendency for clogging or adhesion of suspended
particles~\cite{ybert.c:2007,vinogradova.oi:2011,rothstein.jp:2010}. The effective hydrodynamic slip of anisotropic SH surfaces is generally tensorial~\cite{bazant2008,feuillebois.f:2009,schmieschek.s:2012}, due to secondary flows transverse to the direction of the applied pressure gradient~\cite{feuillebois.f:2010b}. The transverse viscous flow generated by anisotropic SH surfaces could be used to guide the controlled lateral displacement of particles and their separation by size, but these effects remain largely unexplored. Some concepts of fractionation in SH channels of low Re are known. However, their use requires a sedimentation of   particles~\cite{pimponi.d:2014,asmolov_LabChip}, so that they cannot be used for neutrally-buoyant microparticles. In principle, SH channels can be designed to achieve inertial separations of neutrally-buoyant particles, but
we are unaware of any prior work.

\begin{figure}[h!]
\centering
  \includegraphics[width=0.5\textwidth]{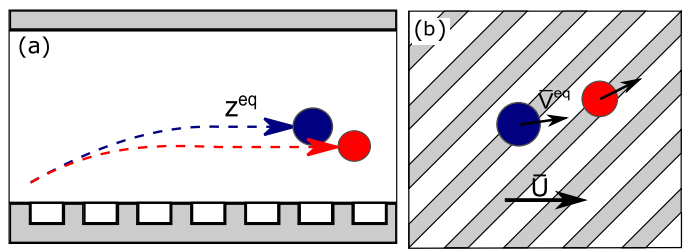}
    \caption{Sketch of the system: side (a) and top (b) views, with a
schematic of vertical and transverse migration. }
  \label{sketch}
\end{figure}

In this paper, we study an inertial migration of neutrally-buoyant spherical particles in a plane channel with the bottom wall decorated by SH grooves, tilted relative to a pressure gradient (as shown in Fig.~\ref{sketch}). Our consideration
is based on the theoretical analysis and lattice-Boltzmann simulations. We show that the effective hydrodynamic slippage of the bottom SH surface affects the migration velocity of particles and alters their equilibrium positions in the channel. We also demonstrate that
particles migrate in the transverse direction with a velocity that depends on
their size. Our results may guide the design of inertial microfluidic devices for efficient sorting of neutrally-buoyant microparticles by their size.

The structure of our manuscript is as follows. In Sec.~\ref{sec:theory}, we define our system and present some theoretical estimates of
particle migration in a channel with a SH wall. Sec.~\ref{sec:Simulation}  discusses our simulation method and justifies the choice of parameters.  Results are discussed in Sec.~\ref{sec:Results} and we conclude in
Sec.~\ref{sec:Conclusion}.

\section{Model and theoretical estimates}
\label{sec:theory}

\begin{figure}[t!]
\centering
  \includegraphics[width=8cm]{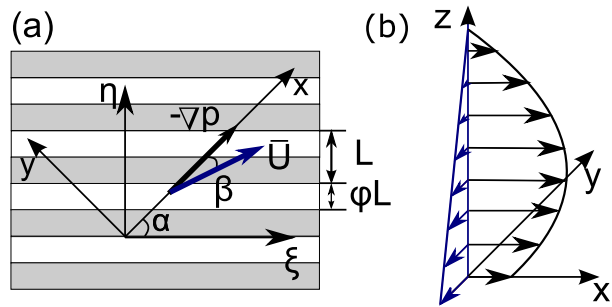}
    \caption{(a) Sketch of a striped superhydrophobic surface with coordinate systems aligned along $-\nabla p$ and stripes; (b) A parabolic forward and a linear transverse shear flows in a SH channel.}
  \label{coords}
\end{figure}

We first present the model and basic physical ideas
underlying our work. We consider a viscous fluid confined between a SH
plate located at $z=0$ and a no-slip hydrophilic plate located at $z=H$ and unbounded in the $x$ and $y$ directions as sketched in Fig.~\ref{coords}(a). The flow is driven by pressure gradient, $\nabla p$, applied at some angle $\alpha$
to the direction of the SH grooves. The $x$-axis is defined along $-\nabla p$, and the $y$-axis is aligned across the channel.
The SH surface is modeled as a flat interface with alternating no-slip and
perfect-slip stripes of a period $L$ and a slipping area fraction $\phi$. In this idealization, we have neglected an additional mechanism for a dissipation connected with the meniscus curvature~\cite{sbragaglia.m:2007,Teo&Khoo:09,nizkaya.tv:2018} which may have an influence on a channel flow.

Our final aim is to calculate translation of particles of radius $a$ in the $x$ and $y$ directions, and their migration in the  direction $z$. To evaluate these we have to know the properties of an undisturbed by particles flow field. We, therefore, first summarize and give some new interpretation of earlier results on fluid velocities in the SH channel.  The maximum velocity of the Poiseuille flow in a no-slip (hydrophilic)
channel of the same thickness would be $U'_m = |\nabla p| H^2 / (8 \mu)$, where $\mu$ is the dynamic viscosity. In our analysis below we use dimensionless, scaled by this value, velocities, and the channel Reynolds
number is then defined as $\Re = \rho U'_{m}H/\mu $, where $\rho $ is the density of a fluid (and particles).

Since the Navier-Stokes equations in the SH channel can be solved only numerically, it is instructive to  analyze the velocity profiles
$\overline{\mathbf{U}}=(\overline{U}_x,\overline{U}_y)$, averaged over the
texture period. When $\Re \ll 1$, the averaged velocity in the channel satisfies tensorial slip boundary
conditions at the SH wall, ${\mathbf{U}}_s=-\mathbf{b}\cdot\partial
\overline{\mathbf{U}}(0)/\partial z$, where ${\mathbf{U}}_s=(U_{sx},U_{sy})$ is
the slip velocity at the SH surface and $\mathbf{b}$ is the effective slip length
tensor. Its eigenvalues $b_{\parallel}$ and $b_{\perp}$ for the striped
texture correspond to the fastest (greatest forward slip) and slowest (least
forward slip) directions, along and across the stripes respectively, and depend
on $\phi$ and $L/H$~\cite{schmieschek.s:2012}. Let us now assume that at finite $\Re$, the tensorial slip approach could serve as a  rough approximation. If so, 
the averaged forward velocity profile would be approximately
parabolic (see Fig.~\ref{coords}(b))
\begin{equation}
  \overline{U}_{x}\simeq 4\left( 1-z/H\right)z/H+U_{sx}\left( 1-z/H\right),
\label{Uxi}
\end{equation}
and an averaged velocity of the transverse shear flow is given by
\begin{equation}
  \overline{U}_{y}\simeq U_{sy}\left(1-z/H\right).
\label{Ueta}
\end{equation}
Averaged forward and transverse slip velocities can be evaluated in terms of
the effective slip tensor~\cite{bazant2008}:
\begin{equation}
U_{sx}\simeq \frac{4b_{x}}{H} ,\quad
U_{sy}\simeq \frac{4b_{y}}{H} ,
\label{usxy}
\end{equation}
where the forward and transverse slip lengths are
\begin{eqnarray}
   b_{x}&\simeq&b_{\parallel}\cos ^{2}\alpha +b_{\perp}\sin
^{2}\alpha,
\label{b_eff_x_thick}
\\ b_{y}&\simeq&\frac{\sin (2\alpha)}{2} \left( b_{\parallel}
-b_{\perp}\right) .
\label{b_eff_xy_thick}
\end{eqnarray}
The angles $\alpha = 0$ and $90^{\circ}$ correspond then to special (eigen) directions,
where $b_x = b_{\parallel}$ and $b_{\perp}$ along which a secondary transverse flow is not generated, $b_y=0$. The maximum transverse flow is generated at
$\alpha = 45^{\circ}$~\cite{vinogradova.oi:2011}.

It is now convenient to define a second coordinate system with the $\xi$-axis directed along the stripes, and
$\eta$ across them (see Fig.~\ref{coords}(a)). The reason is an undisturbed flow field, which depends only
on $\eta$ and $z$, is periodic in the $\eta$-direction. Besides, as we report in Sec.~\ref{sec:Simulation}, our simulation cell is also periodic in the $\xi$ direction, so that all simulation variables will be averaged over these two coordinates. In this coordinate system, the averaged velocity
field is given by
\begin{eqnarray}
  \overline{U}_{\xi}&=&\overline{U}_{x}\cos \alpha -\overline{U}_{y}\sin \alpha , \\
 \overline{U}_{\eta}&=&\overline{U}_{x}\sin \alpha +\overline{U}_{y}\cos \alpha .
 \label{Uxieta}
\end{eqnarray}
At a finite channel Reynolds numbers, $\Re \geq 1$, the averaged velocity profile can
generally differ from the parabolic one due to the presence of inertial terms in the Navier-Stokes
equations. However, as we will see below in Sec.~\ref{sec:Simulation}, in reality, the deviations
from the creeping-flow limit, Eqs.~(\ref{Uxi})-(\ref{b_eff_xy_thick}), are negligibly small even when Re$ \simeq 20$.

We now turn to the motion of (force and torque free) neutrally-buoyant
particles.  In an unbounded linear shear flow such particles
would move with the velocity $\mathbf V$, which is equal to the fluid velocity $\mathbf U$ at
the particle's position $z$. In the case of an unbounded parabolic flow, they would slightly
lag the fluid due to Faxen corrections. However, the particle velocity in the channel
can differ significantly from that of a fluid due to hydrodynamic interactions
with the walls. It has recently been suggested that in a symmetric no-slip hydrophilic channel this difference can be characterized by the correction functions
$h_{\xi}$ and $h_{\eta}$~\cite{asmolov2018}:
\begin{equation}
\overline{V}_{\xi}=h_{\xi} (z/a) \overline{U}_{\xi} (z), \quad
\overline{V}_{\eta}= h_{\eta} (z/a)  \overline{U}_{\eta} (z).
  \label{v_xi}
\end{equation}
These two functions, $0\leq h_{\xi,\eta}\leq 1$, should obviously become different for our SH channel due to the wall slippage and anisotropy, as well as an asymmetry of the channel itself. Note that, as shown before~\cite{davis1994,loussaief15,ghalia2016}, near a homogeneous slippery wall a particle has a smaller lag, $\mathbf {U - V}$, compared to the no-slip case, so that functions $h_{\xi,\eta}$ become close to unity. This lag should be even smaller for highly slippery  SH wall.

An inertial lift force at finite Re is directed normal to the channel walls, and scales
as~\cite{Asmolov99}
\begin{equation}
F_{l}=\rho (4a^{2}U_m^{\prime}/H)^2c_{l},
\label{cher}
\end{equation}%
where $c_{l}(z/H,a/H,\Re)$ is the lift coefficient. The lift in a symmetric no-slip
channel is an antisymmetric function of $z/H-1/2$ due to the flow symmetry. This function
has three zeros, which correspond to equilibrium
positions of particles. Two of them are stable, being symmetric and located at the vicinity of the walls. The third zero has a locus at the midplane of the channel, where the shear rate
$dU/dz$ vanishes, and corresponds to an unstable equilibrium. In a SH channel the secondary transverse flow is usually much weaker than the forward flow, so that $|\overline{U}_{x}|\gg|\overline{U}_{y}|$. We, therefore, could expect that the lift  force will be generated
mainly by the gradient of the forward velocity, similarly to a hydrophilic channel, but the locations $\zeq$ of three equilibrium positions should be, of course, altered due to a bottom wall slip. Since the difference $\mathbf {U - V}$ is expected to become smaller in the presence of the SH wall, a weaker lift force could be expected as compared with reported for a hydrophilic channel~\cite{asmolov2018}.

Finally, we expect that the transverse fluid velocity induces a lateral translation of particles. Since their equilibrium
positions depend on radii, the particles of different sizes are expected to move with different transverse velocities
$\overline{V}_y^{\eq}$, allowing for their spatial separation in the $y-$
direction.

\section{Simulation method}

\label{sec:Simulation}

Fluid flow and particle motion in the  channel are simulated using the lattice
Boltzmann method~\citep{benzi_lattice_1992,kunert2010random}. We use a 3D, 19
velocity, single relaxation time implementation of the lattice Boltzmann method
with a Batnagar Gross Krook (BGK) collision operator. The relaxation time
$\tau$ is equal to $1$ throughout this paper, so that the kinematic viscosity
in simulation units is $\nu=1/6$. The fluid density in the initial state
$\rho_0$ is set to 1 and remains approximately constant throughout the
simulations. Spherical particles with radius $a$ are implemented as moving
no-slip boundaries originally proposed by Ladd~\citep{LaddVerberg2001}. Our
implementation has been used before~\cite{kunert2010random,bib:jens-janoschek-toschi-2010b,JHT14a,DSAHV14,HFRRWL14}
including a recent related article on a particle experiencing an inertial lift
force in a no-slip channel~\cite{asmolov2018}.

The simulation domain has dimensions $N_{\xi}=N_{\eta}=128$ and $N_z=81$
lattice cells, which corresponds to a channel of height $H=80$. Periodic
boundary conditions are applied on the sides. A mid-grid bounce back no-slip
boundary condition is implemented at the top wall and an on-site slip boundary
condition is applied to the bottom wall to describe the striped
texture~\cite{AhmedHecht2009,HechtHarting2010,schmieschek.s:2012}. The stripes are parallel to the
$\xi$-axis and have widths $\phi L$ (perfect slip)  and $(1-\phi)L$ (no slip)
with $L=32$ ($L/H=0.4$) and $\phi$ varying from $0$ (no-slip) to $0.875$. Since
it is impossible to implement the pressure gradient explicitly in a periodic
setup, we model it by applying a volumetric body force to the fluid. An
equivalent force is applied to particles to ensure neutral buoyancy. The body
force enters the lattice Boltzmann algorithm by shifting the equilibrium
velocity in the collision operator (Shan-Chen forcing). It is applied at an
angle $\alpha=45^\circ$ to the stripes to maximize a transverse shear, and its magnitude is set to $g=10^{-5}$
in simulation units. In a channel with no-slip walls this force creates a
Poiseuille flow with $\Re \simeq 23$.

Particles are released from different initial positions with a velocity equal
to the average flow velocity at the corresponding height. Simulations run until
the particle reaches its equilibrium position in $z$. Typically, this takes
about $300000$ time steps. The equilibrium velocities $\overline{V}_x^{\eq}$
and $\overline{V}_y^{\eq}$ are obtained by averaging over an interval of
$\Delta t=1000$ time steps.

\section{Results and discussion}
\label{sec:Results}

In this section, we present our simulation results. More specifically, we discuss simulation data obtained for vertical and lateral displacement of particles, and also present some data on a fluid velocity field and slippage at the SH wall.

\begin{figure}[h!]
\centering
 \includegraphics[width=0.6\textwidth]{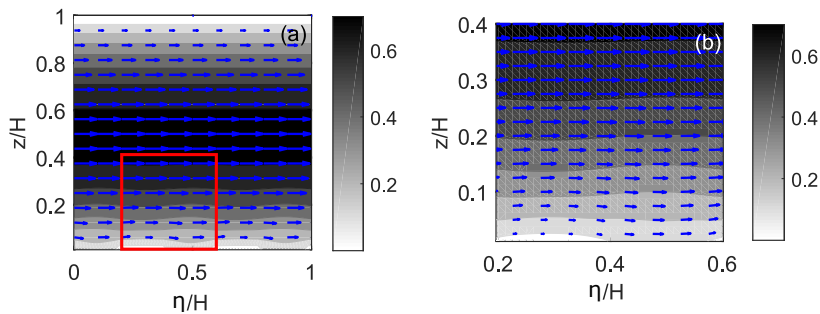}
    \caption{(a) Cross-section of the velocity field ($U_{\eta}$, $U_z$) computed at $\phi=0.5$. Color map shows the velocity component $U_{\xi}$. (b) A zoom in of the  marked by a square region.}
  \label{VelField}
\end{figure}

\begin{figure}[h!]
\centering
 \includegraphics[width=0.6\textwidth]{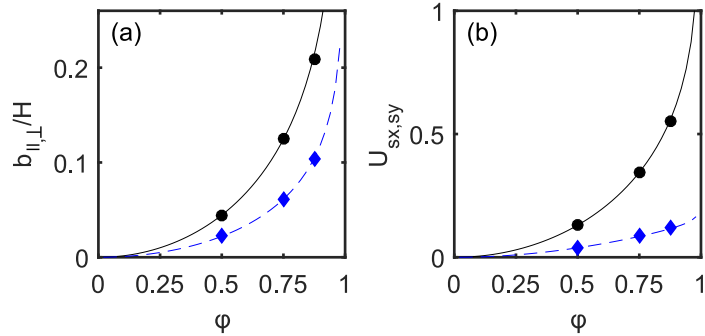}
    \caption{(a) Longitudinal  (circles) and transverse (diamonds) effective slip lengths obtained in simulations. Solid and dashed curves show results of theoretical calculations. (b) Corresponding slip velocities.}
  \label{beff}
\end{figure}

It is by no means obvious that at Re$\simeq 20$ the approximate equations (\ref{Uxi})-(\ref{b_eff_xy_thick}) will be accurate enough. We, therefore, first simulate the flow in a particle-free channel to characterize the
unperturbed velocity field and to check the validity of formulas. A typical velocity field for a channel with a SH bottom
wall is shown in Fig.~\ref{VelField}. The fluid has a finite velocity at the
slippery parts of the texture. The inhomogeneous boundary conditions result in flow
ondulations near the bottom wall, but at distances comparable to $L$ the flow
becomes homogeneous in the $\eta$ direction.
We obtain the averaged velocity profiles in the directions along and across
stripes, $\overline{U}_\xi(z)$ and $\overline{U}_\eta(z)$, by integrating the
velocity field over the texture period in $\eta$. The eigenvalues of effective
slip lengths are then calculated as
\begin{equation}\label{sliplength}
    b_{\|,\perp}=\left.\dfrac{\overline{U}_{\xi,\eta}(z)}{d\overline{U}_{\xi,\eta}/dz}\right|_{z=0}.
\end{equation}
Fig.~\ref{beff}(a) shows the effective slip lengths $b_{\|,\perp}$ at the SH wall as a function
of $\phi$. Also included are theoretical curves obtained using the Fourier
method~\cite{nizkaya2013} and valid for $\Re \ll 1$. We can see that the fits of our finite Re simulation data are
very good. These data and theoretical calculations are used to calculate the forward and transverse slip velocities from Eqs.~(\ref{usxy})-(\ref{b_eff_xy_thick}). The results are shown in Fig.\ref{beff}(b), and we again observe that theoretical and simulation data nearly coincide. We can, therefore, conclude that calculations based on a low Re theory provide a good description of the slip lengths and averaged slip velocities obtained in simulations.

\begin{figure}[h!]
\centering
 \includegraphics[width=0.6\textwidth]{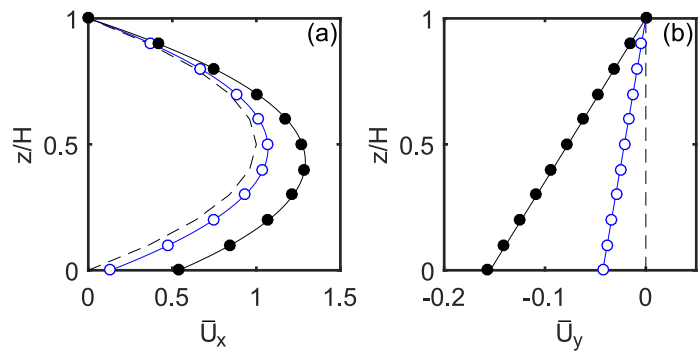}
    \caption{Averaged velocity profiles along (a) $x-$ and (b) $y-$ axes.
Symbols show simulation results obtained using $\phi=0.5$ (open circles) and $\phi=0.875$
(filled circles). Theoretical predictions are shown by solid curves. Dashed curves plot theoretical expectations for a hydrophilic channel of $\phi=0$.}
  \label{prof}
\end{figure}

Fig.~\ref{prof} plots the averaged velocity profiles in the directions $x$ and $y$. The simulations are made using  $\phi=0.5$ and $0.875$. We can see that the forward velocity profiles are always parabolic, and that the  shear rate of the secondary transverse flow is uniform
throughout the channel. For similar values of $\Re$, but for  a flow between two  misaligned striped SH walls,  it has earlier been found that the creeping-flow equations are violated, and the transverse flow is suppressed by the fluid inertia~\cite{nizkaya2015}. We have compared our simulation results with theoretical calculations from Eqs.~(\ref{Uxi}) and (\ref{Ueta}) and see that in our case of a channel with one SH wall, the fluid inertia does not alter the $ \overline{U}_{x}$ and $ \overline{U}_{y}$ profiles. Theoretical curves for a hydrophilic channel ($\phi=0$)
are also shown in Fig.~\ref{prof}. In the hydrophilic channel the flow is symmetric and the transverse shear is not generated. It can be seen that with the increase in fraction of the slipping area $\phi$, the absolute values of forward and transverse velocities grow, and the $ \overline{U}_{x}$ profile becomes asymmetric with the displacement of its maximum towards a SH wall.

\begin{figure}[h]
\centering
  \includegraphics[width=0.53\textwidth]{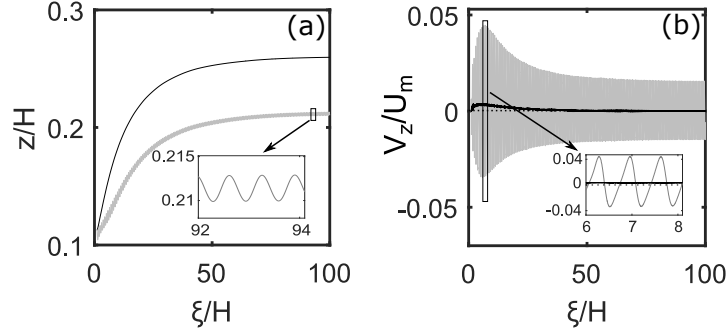}
    \caption{(a) Trajectories and (b) normal velocities of the particle of
$a/H=0.1$ in the hydrophilic channel (black curves) and the channel with a SH wall of
$\phi=0.875$ (gray curves). }
  \label{trajs}
\end{figure}

\begin{figure}[h]
\centering
  \includegraphics[width=0.6\textwidth]{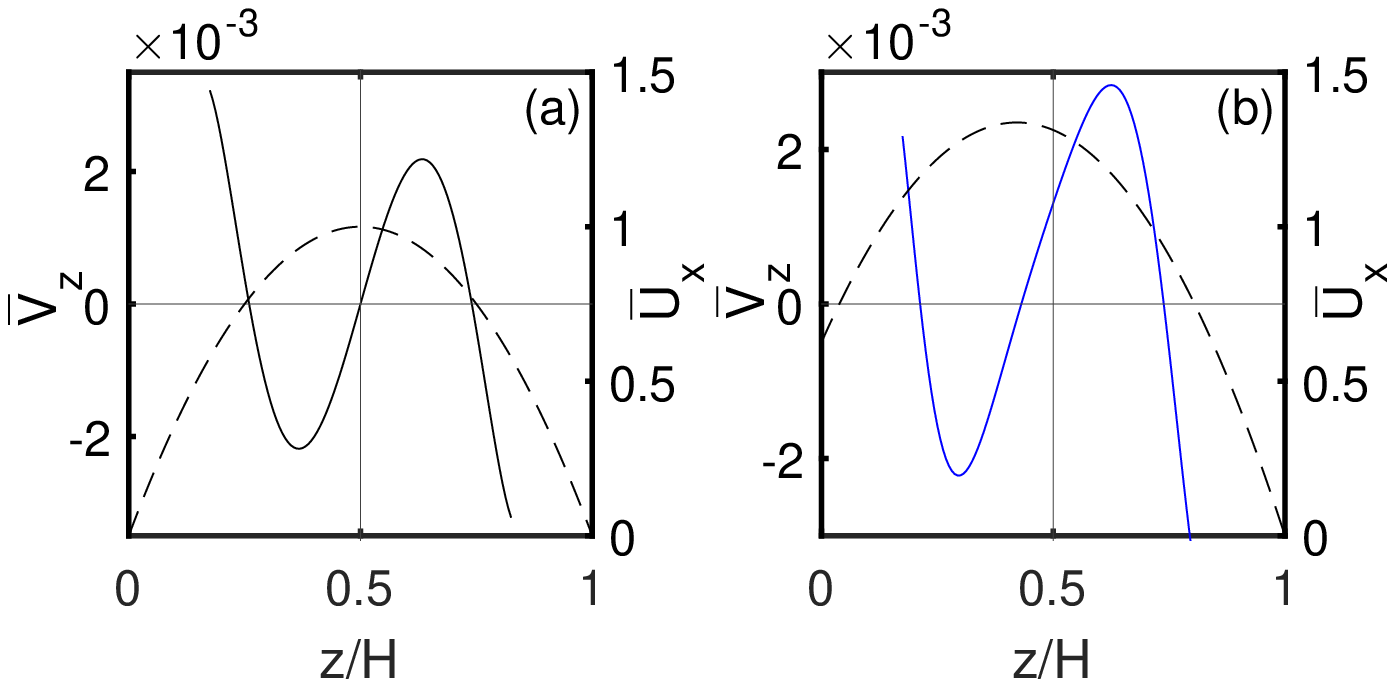}
    \caption{Migration velocity of a particle of radius $a/H=0.1$ as a function of a cross-stream particle position (solid curves) in (a) the hydrophilic no-slip channel and (b) the channel with the bottom SH wall of $\phi=0.875$. Dashed curves plot forward velocity profiles.}
  \label{migr}
\end{figure}


As demonstrated above (see Figs.~\ref{beff} and ~\ref{prof}), the slip velocity at the bottom wall $U_{s x}$ and $U_{sy}$ increases with
$\phi$  and the transverse velocity $ \overline{U}_{y}$ is largest at the
bottom SH wall and decreases linearly to zero at the top wall.
 Therefore, it is natural to expect that focusing at different $\zeq$ neutrally buoyant particles, will translate in the transverse direction with different velocities. We now launch particles at some arbitrary $z$ and with an initial velocity equal to the averaged liquid
velocity at this height, and monitor particle trajectories. This allows one to determine the particle migration velocity $\overline{V}_z$ as a function of $z$. Since  $\overline{V}_z$ and $F_l$ vanish at the same $z$, the zeros of $\overline{V}_z$ correspond to the focusing positions of particles.
 Fig.~\ref{trajs}(a) shows trajectories of a particle
of $a/H=0.1$ launched in the vicinity of the bottom hydrophilic and SH walls. In the case of hydrophilic no-slip wall, the particle reaches its equilibrium position after translating to a distance, roughly equal to
$100$ channel width. At larger distances its $z$-position remains nearly
constant, and $V_z$ becomes negligibly small. When we deal with a SH bottom wall,  both the
particle cross-stream coordinate and the velocity oscillate about some mean values. Note that the
amplitude of the velocity oscillations (shown in  Fig.~\ref{trajs}(b)) is large compared with the migration
velocity for the no-slip case. This indicates that these oscillations are not
due to the lift force, and are induced by the variation of the $z-$component of the fluid
velocity (see Fig.~\ref{VelField}). The oscillations
in the particle position are, however, relatively small. The average particle trajectory is qualitatively similar to observed in the hydrophilic channel, but the equilibrium  position is much closer to the SH wall. This is likely due to a weaker hydrodynamic interaction with a slippery wall~\cite{DSAHV14}.

To obtain the dependence of the migration velocity $\overline{V}_z(z)$ on $z$ we launch particles of radius $a/H=0.1$ from different initial
positions and, after computation of trajectories and averaging out the oscillations, obtain the results
presented in Fig.~\ref{migr}. We  recall that $\overline{V}_z$ vanishes at equilibrium positions $\zeq$. In the no-slip channel (Fig.~\ref{migr}(a)), the migration velocity curve is similar to reported earlier for smaller particles~\cite{asmolov2018}. The flow in the channel is symmetric, leading to focusing of particles at two symmetric equilibrium positions. There is also an
unstable equilibrium position located at the channel midplane. In the SH channel (Fig.~\ref{migr}(b)) the fluid forward flow is asymmetric and faster, with the maximum of $\overline{U}_x$ shifted towards the slippery wall. This modifies the profile of $\overline{U}_x (z)$ compared with a hydrophilic channel. There are still two stable and one unstable equilibrium positions, but it is well seen that they are shifted towards the slippery wall. We remark that $\zeq$ of an unstable position roughly coincides with the coordinate of maximum of $\overline{U}_x$.

\begin{figure}[h]
\centering
  \includegraphics[width=0.6\textwidth]{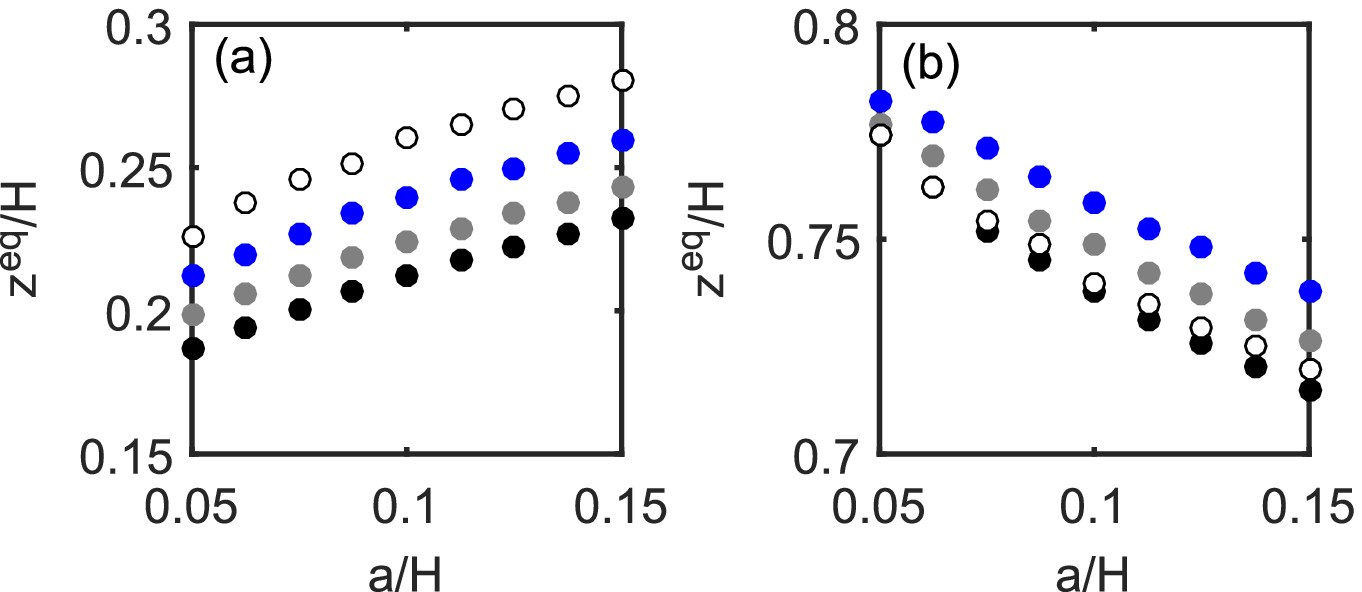}
    \caption{(a) Lower and (b) upper equilibrium positions vs. particle radius. Open circles show results for a no-slip channel, $\phi=0$. Filled symbols from top to bottom plot simulation results obtained using $\phi = 0.5$, $0.75$, and $0.875$.}
  \label{z_eq}
\end{figure}

It is known that focusing positions of particles in a hydrophilic channel depend, although slightly, on their
size~\cite{asmolov2018}. We now investigate the effect of particle radii on two stable $\zeq$ in the presence of a SH wall, by varying
 $a/H$ from $0.05$ to $0.15$ and by using SH textures of $\phi=0.5$, $0.75$ and $0.875$. The
simulation results are shown in Fig.~\ref{z_eq} together with data for a hydrophilic channel. It is seen in Fig.~\ref{z_eq}(a) that the lower $\zeq$ increases with the particle size, but decreases with $\phi$, i.e. with the amplitude of the wall slip, as predicted in Sec.~\ref{sec:theory}. The distance of the second stable focusing position from the upper wall also increases with the particle size (see Fig.~\ref{z_eq}(b)). An unexpected  result, however, is that
these upper equilibrium positions are also significantly
affected by the slip at the bottom wall. One can see deviations from the no-slip case even with a rather small slipping fraction, $\phi=0.5$, when $b_x/H$ is relatively small, below 0.05. Another startling conclusion from this plot is that the dependence of the upper $\zeq$ on $\phi$ is non-monotonous, and the equilibrium distance
to the upper wall takes its minimum at $\phi=0.5$.

\begin{figure}[h!]
\centering
  \includegraphics[width=0.6\textwidth]{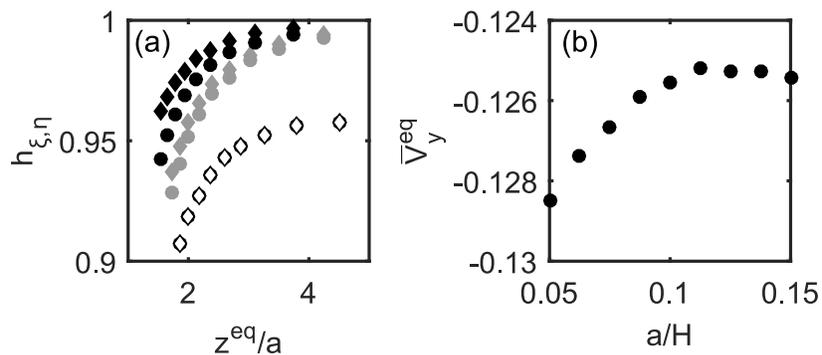}
    \caption{(a) Correction functions $h_{\xi}$ (filled diamonds) and $h_{\eta}$
(filled circles) for particles at the lower equilibrium positions computed for $\phi = 0.5$ (gray) and
$0.875$ (black). Open symbols show results obtained using $\phi = 0$; (b) lateral
$\overline{V}_{y}^{\eq}$ particle velocities at the lower equilibrium positions computed
for $\phi=0.875$.}
  \label{V_eq}
\end{figure}

At equilibrium particles translate at a plane $z=z^{\eq}$
with a mean velocity $(\overline{V}^{\eq}_x$, $\overline{V}^{\eq}_y)$. This velocity
can be obtained from simulation data. Then, the correction functions $h_{\xi,\eta}$ in
Eq.~(\ref{v_xi}), which characterize the difference between the average particle
and fluid velocities, can be found.
The results for lower equilibrium positions are presented
in Fig.~\ref{V_eq}(a). The simulation data shows the correction functions grow with $\zeq /a$, and
for SH channels are well above the correction $h$ for a hydrophilic channel. The deviations from $h$ increases with $\phi$. We also see that
$h_{\xi}$ is larger than $h_{\eta}$, especially for the largest $\phi=0.875$ used in simulations.
Fig.~\ref{V_eq}(b) shows the transverse component of the equilibrium
velocity as a function of the particle size computed for a SH channel of $\phi=0.875$. We have initially assumed that this should increase with the particle size. It is seen, however, that $\overline{V}^{\eq}_y(a/H)$ saturates when
$a/H \geq 0.1$.

\begin{figure}[h]
\centering
\includegraphics[width=0.6\textwidth]{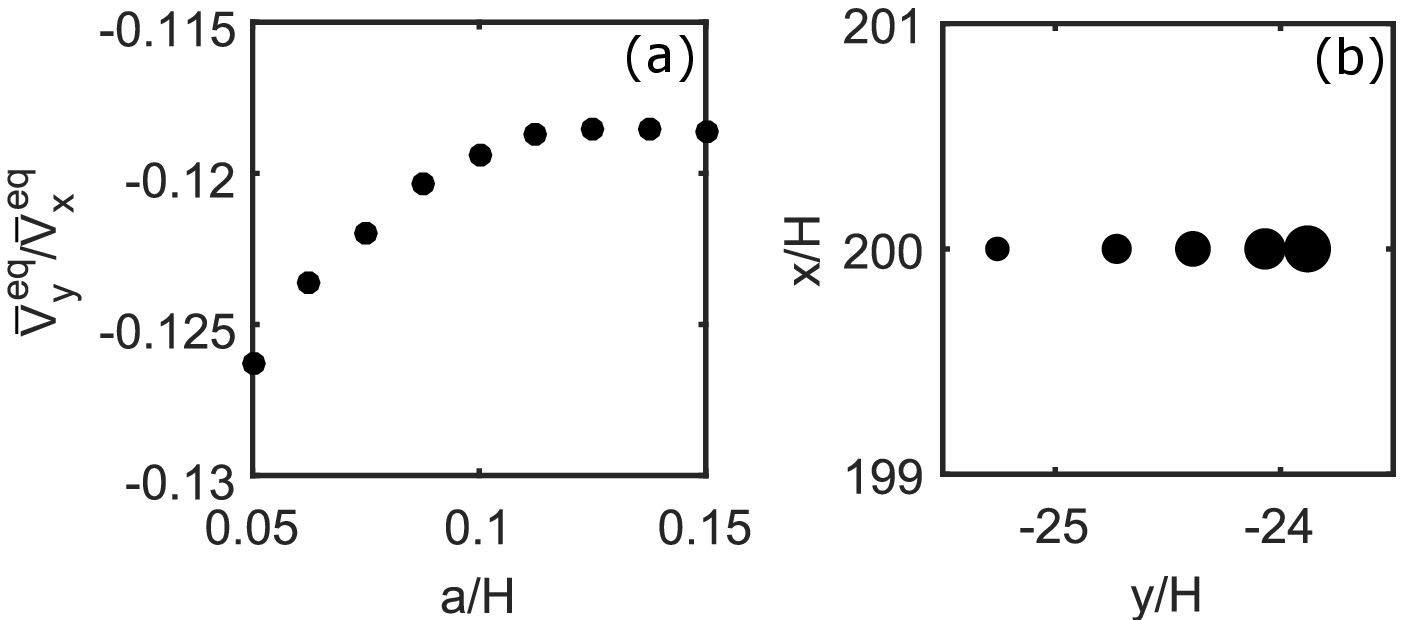}
    \caption{(a) The ratio of forward and transverse equilibrium velocities as a function of
particle size; (b) Lateral positions of particles at the outlet of the SH channel. Simulations are made for particles of radius from
$a/H=0.05$ to $0.1$ with a step $0.0125$ (from left to right).}
  \label{outlet}
\end{figure}

Finally, we investigate the lateral displacements of particles, which are controlled by the ratio
of their forward and transverse velocities,
$\overline{V}^{\eq}_y/\overline{V}^{\eq}_x$. The dependence of this ratio on $a / H$ is shown in Fig.~\ref{outlet}(a). We see that it increases when $a/H \leq 0.1$, but saturates for larger particles, indicating that particles of $a/H \geq 0.1$ cannot be laterally separated. We now launch the particles at $z^{\eq}$, and measure their lateral displacement at the outlet of the channel of length $l=200H$. The simulation results are shown in Fig.~\ref{outlet}(b), where the size of symbols reflects the particle radius. This plot indicates that small particles are well separated by distances of the order of the channel height, which is well above the distance between them in the $z$-direction (cf. Fig.~\ref{z_eq}), and
several times greater than the particle size itself. Separation can, of course, be further
improved by increasing the channel length.

%

\section{Conclusion}
\label{sec:Conclusion}
We have investigated the inertial migration of neutrally-buoyant particles in a plane channel with a bottom SH striped wall tilted relative to a pressure gradient. It has been shown that the flow near SH stripes is inhomogeneous and periodic, and leading to regular oscillations of the
particle normal velocity and position. We have found that the SH slip significantly affects the lift force and alters all three focusing positions of
neutrally buoyant particles in the channel. The lower (stable) focusing position has been shown to be significantly shifted towards
the SH wall due to weaker hydrodynamic interactions. We have also demonstrated that the second stable
focusing position (near upper hydrophilic wall) is also affected by the SH wall slip, and that the shift in its location is discernible even at a relatively small slipping fraction at the bottom wall, when its effective slip is moderate.

Our results show that a transverse shear flow generated by inclined SH
stripes leads to a particle translation in the direction transverse to the mean (forward) flow, and that the lateral speed of small particles ($a/H \leq 0.1$) increases with their size. The lateral distance between particles of different radii grows
with channel length, suggesting that this can be used for their separation.
It follows from our study that to separate particles by a distance of a few diameters,
the channel length-to-height ratio should be around 200. For example, to separate laterally the neutrally-buoyant particles of radii $3\,\mu\mathrm{m}$ and $4\,\mu\mathrm{m}$ in a channel of height $40\,\mu\mathrm{m}$, the channel
length should be about $8\,\mathrm{mm}$. However, we can conclude that it would be impossible to employ our strategy for a fractionation of larger particles.

Our work can   be   extended   to a lateral separation of heavy particles, and one can expect that in this case the proposed concept may be even more efficient. Heavy particles may be separated not only by size, but also a density. Such particles should focus closer to the bottom wall, where the transverse shear rate is maximal, and where the difference between forward velocities of the particle and of the undisturbed flow, which strongly depends on the particle size, increases drastically. All these could potentially lead to an efficient lateral separation.

\begin{acknowledgments}
This research was partly supported by the Russian Foundation for Basic Research (grant 18-01-00729), the Ministry of Science and Higher Education of the Russian Federation and by the German Research Foundation (grant HA 4382/8-1).
\end{acknowledgments}

\appendix

\bibliography{lift_new}

\begin{thebibliography}{42}%
\makeatletter
\providecommand \@ifxundefined [1]{%
 \@ifx{#1\undefined}
}%
\providecommand \@ifnum [1]{%
 \ifnum #1\expandafter \@firstoftwo
 \else \expandafter \@secondoftwo
 \fi
}%
\providecommand \@ifx [1]{%
 \ifx #1\expandafter \@firstoftwo
 \else \expandafter \@secondoftwo
 \fi
}%
\providecommand \natexlab [1]{#1}%
\providecommand \enquote  [1]{``#1''}%
\providecommand \bibnamefont  [1]{#1}%
\providecommand \bibfnamefont [1]{#1}%
\providecommand \citenamefont [1]{#1}%
\providecommand \href@noop [0]{\@secondoftwo}%
\providecommand \href [0]{\begingroup \@sanitize@url \@href}%
\providecommand \@href[1]{\@@startlink{#1}\@@href}%
\providecommand \@@href[1]{\endgroup#1\@@endlink}%
\providecommand \@sanitize@url [0]{\catcode `\\12\catcode `\$12\catcode
  `\&12\catcode `\#12\catcode `\^12\catcode `\_12\catcode `\%12\relax}%
\providecommand \@@startlink[1]{}%
\providecommand \@@endlink[0]{}%
\providecommand \url  [0]{\begingroup\@sanitize@url \@url }%
\providecommand \@url [1]{\endgroup\@href {#1}{\urlprefix }}%
\providecommand \urlprefix  [0]{URL }%
\providecommand \Eprint [0]{\href }%
\providecommand \doibase [0]{http://dx.doi.org/}%
\providecommand \selectlanguage [0]{\@gobble}%
\providecommand \bibinfo  [0]{\@secondoftwo}%
\providecommand \bibfield  [0]{\@secondoftwo}%
\providecommand \translation [1]{[#1]}%
\providecommand \BibitemOpen [0]{}%
\providecommand \bibitemStop [0]{}%
\providecommand \bibitemNoStop [0]{.\EOS\space}%
\providecommand \EOS [0]{\spacefactor3000\relax}%
\providecommand \BibitemShut  [1]{\csname bibitem#1\endcsname}%
\let\auto@bib@innerbib\@empty
\bibitem [{\citenamefont {Di~Carlo}\ \emph {et~al.}(2007)\citenamefont
  {Di~Carlo}, \citenamefont {Irimia}, \citenamefont {Tompkins},\ and\
  \citenamefont {Toner}}]{dicarlo07}%
  \BibitemOpen
  \bibfield  {author} {\bibinfo {author} {\bibfnamefont {D.}~\bibnamefont
  {Di~Carlo}}, \bibinfo {author} {\bibfnamefont {D.}~\bibnamefont {Irimia}},
  \bibinfo {author} {\bibfnamefont {R.G.}\ \bibnamefont {Tompkins}}, \ and\
  \bibinfo {author} {\bibfnamefont {M.}~\bibnamefont {Toner}},\ }\bibfield
  {title} {\enquote {\bibinfo {title} {Continuous inertial focusing, ordering,
  and separation of particles in microchannels},}\ }\href@noop {} {\bibfield
  {journal} {\bibinfo  {journal} {Proc. Natl. Acad. Sci. USA}\ }\textbf
  {\bibinfo {volume} {104}},\ \bibinfo {pages} {18892--18897} (\bibinfo {year}
  {2007})}\BibitemShut {NoStop}%
\bibitem [{\citenamefont {Bhagat}\ \emph {et~al.}(2008)\citenamefont {Bhagat},
  \citenamefont {Kuntaegowdanahalli},\ and\ \citenamefont
  {Papautsky}}]{bhagat08}%
  \BibitemOpen
  \bibfield  {author} {\bibinfo {author} {\bibfnamefont {A.~A.~S.}\
  \bibnamefont {Bhagat}}, \bibinfo {author} {\bibfnamefont {S.~S.}\
  \bibnamefont {Kuntaegowdanahalli}}, \ and\ \bibinfo {author} {\bibfnamefont
  {I.}~\bibnamefont {Papautsky}},\ }\bibfield  {title} {\enquote {\bibinfo
  {title} {Continuous particle separation in spiral microchannels using dean
  flows and differential migration},}\ }\href@noop {} {\bibfield  {journal}
  {\bibinfo  {journal} {Lab. Chip}\ }\textbf {\bibinfo {volume} {8}},\ \bibinfo
  {pages} {1906--1914} (\bibinfo {year} {2008})}\BibitemShut {NoStop}%
\bibitem [{\citenamefont {Gossett}\ \emph {et~al.}(2010)\citenamefont
  {Gossett}, \citenamefont {Weaver}, \citenamefont {Mach}, \citenamefont {Hur},
  \citenamefont {Tse}, \citenamefont {Lee}, \citenamefont {Amini},\ and\
  \citenamefont {Di~Carlo}}]{gossett2010}%
  \BibitemOpen
  \bibfield  {author} {\bibinfo {author} {\bibfnamefont {D.~R.}\ \bibnamefont
  {Gossett}}, \bibinfo {author} {\bibfnamefont {W.~M.}\ \bibnamefont {Weaver}},
  \bibinfo {author} {\bibfnamefont {A.~J.}\ \bibnamefont {Mach}}, \bibinfo
  {author} {\bibfnamefont {S.~C.}\ \bibnamefont {Hur}}, \bibinfo {author}
  {\bibfnamefont {H.~T.~K.}\ \bibnamefont {Tse}}, \bibinfo {author}
  {\bibfnamefont {W.}~\bibnamefont {Lee}}, \bibinfo {author} {\bibfnamefont
  {H.}~\bibnamefont {Amini}}, \ and\ \bibinfo {author} {\bibfnamefont
  {D.}~\bibnamefont {Di~Carlo}},\ }\bibfield  {title} {\enquote {\bibinfo
  {title} {Label-free cell separation and sorting in microfluidic systems},}\
  }\href@noop {} {\bibfield  {journal} {\bibinfo  {journal} {Anal. Bioanal.
  Chem.}\ }\textbf {\bibinfo {volume} {397}},\ \bibinfo {pages} {3249--3267}
  (\bibinfo {year} {2010})}\BibitemShut {NoStop}%
\bibitem [{\citenamefont {Giddings}\ \emph {et~al.}(1976)\citenamefont
  {Giddings}, \citenamefont {Yang},\ and\ \citenamefont
  {Myers}}]{Giddings1976}%
  \BibitemOpen
  \bibfield  {author} {\bibinfo {author} {\bibfnamefont {J.~C.}\ \bibnamefont
  {Giddings}}, \bibinfo {author} {\bibfnamefont {F.~J.}\ \bibnamefont {Yang}},
  \ and\ \bibinfo {author} {\bibfnamefont {M.~N.}\ \bibnamefont {Myers}},\
  }\bibfield  {title} {\enquote {\bibinfo {title} {Theoretical and experimental
  characterization of flow field-flow fractionation},}\ }\href@noop {}
  {\bibfield  {journal} {\bibinfo  {journal} {Anal. Chem.}\ }\textbf {\bibinfo
  {volume} {48}},\ \bibinfo {pages} {1126--1132} (\bibinfo {year}
  {1976})}\BibitemShut {NoStop}%
\bibitem [{\citenamefont {Giddings}\ \emph {et~al.}(1977)\citenamefont
  {Giddings}, \citenamefont {Yang},\ and\ \citenamefont
  {Myers}}]{Giddings1977}%
  \BibitemOpen
  \bibfield  {author} {\bibinfo {author} {\bibfnamefont {J.~C.}\ \bibnamefont
  {Giddings}}, \bibinfo {author} {\bibfnamefont {F.~J.}\ \bibnamefont {Yang}},
  \ and\ \bibinfo {author} {\bibfnamefont {M.~N..}\ \bibnamefont {Myers}},\
  }\bibfield  {title} {\enquote {\bibinfo {title} {Correction - theoretical and
  experimental characterization of field-flow fractionation},}\ }\href
  {\doibase 10.1021/ac50011a601} {\bibfield  {journal} {\bibinfo  {journal}
  {Anal. Chem.}\ }\textbf {\bibinfo {volume} {49}},\ \bibinfo {pages}
  {523--523} (\bibinfo {year} {1977})}\BibitemShut {NoStop}%
\bibitem [{\citenamefont {Zhang}\ \emph {et~al.}(1994)\citenamefont {Zhang},
  \citenamefont {Williams}, \citenamefont {Myers},\ and\ \citenamefont
  {Giddings}}]{zhang1994separation}%
  \BibitemOpen
  \bibfield  {author} {\bibinfo {author} {\bibfnamefont {Jue}\ \bibnamefont
  {Zhang}}, \bibinfo {author} {\bibfnamefont {P~Stephen}\ \bibnamefont
  {Williams}}, \bibinfo {author} {\bibfnamefont {Marcus~N}\ \bibnamefont
  {Myers}}, \ and\ \bibinfo {author} {\bibfnamefont {J~Calvin}\ \bibnamefont
  {Giddings}},\ }\bibfield  {title} {\enquote {\bibinfo {title} {Separation of
  cells and cell-sized particles by continuous splitt fractionation using
  hydrodynamic lift forces},}\ }\href@noop {} {\bibfield  {journal} {\bibinfo
  {journal} {Separation science and technology}\ }\textbf {\bibinfo {volume}
  {29}},\ \bibinfo {pages} {2493--2522} (\bibinfo {year} {1994})}\BibitemShut
  {NoStop}%
\bibitem [{\citenamefont {Zhang}\ \emph {et~al.}(2016)\citenamefont {Zhang},
  \citenamefont {Yan}, \citenamefont {Yuan}, \citenamefont {Alici},
  \citenamefont {Nguyen}, \citenamefont {Warkiani},\ and\ \citenamefont
  {Li}}]{zhang2016}%
  \BibitemOpen
  \bibfield  {author} {\bibinfo {author} {\bibfnamefont {J.}~\bibnamefont
  {Zhang}}, \bibinfo {author} {\bibfnamefont {S.}~\bibnamefont {Yan}}, \bibinfo
  {author} {\bibfnamefont {D.}~\bibnamefont {Yuan}}, \bibinfo {author}
  {\bibfnamefont {G.}~\bibnamefont {Alici}}, \bibinfo {author} {\bibfnamefont
  {N.-T.}\ \bibnamefont {Nguyen}}, \bibinfo {author} {\bibfnamefont {M.~E.}\
  \bibnamefont {Warkiani}}, \ and\ \bibinfo {author} {\bibfnamefont
  {W.}~\bibnamefont {Li}},\ }\bibfield  {title} {\enquote {\bibinfo {title}
  {Fundamentals and applications of inertial microfluidics: a review},}\
  }\href@noop {} {\bibfield  {journal} {\bibinfo  {journal} {Lab. Chip}\
  }\textbf {\bibinfo {volume} {16}},\ \bibinfo {pages} {10--34} (\bibinfo
  {year} {2016})}\BibitemShut {NoStop}%
\bibitem [{\citenamefont {Stoecklein}\ and\ \citenamefont
  {Di~Carlo}(2018)}]{stoecklein2018nonlinear}%
  \BibitemOpen
  \bibfield  {author} {\bibinfo {author} {\bibfnamefont {D.}~\bibnamefont
  {Stoecklein}}\ and\ \bibinfo {author} {\bibfnamefont {D.}~\bibnamefont
  {Di~Carlo}},\ }\bibfield  {title} {\enquote {\bibinfo {title} {Nonlinear
  microfluidics},}\ }\href@noop {} {\bibfield  {journal} {\bibinfo  {journal}
  {Anal. Chem.}\ }\textbf {\bibinfo {volume} {91}},\ \bibinfo {pages}
  {296--314} (\bibinfo {year} {2018})}\BibitemShut {NoStop}%
\bibitem [{\citenamefont {Zhang}\ \emph {et~al.}(2014)\citenamefont {Zhang},
  \citenamefont {Yan}, \citenamefont {Alici}, \citenamefont {Nguyen},
  \citenamefont {{Di Carlo}},\ and\ \citenamefont {Li}}]{zhang2014real}%
  \BibitemOpen
  \bibfield  {author} {\bibinfo {author} {\bibfnamefont {J.}~\bibnamefont
  {Zhang}}, \bibinfo {author} {\bibfnamefont {S.}~\bibnamefont {Yan}}, \bibinfo
  {author} {\bibfnamefont {G.}~\bibnamefont {Alici}}, \bibinfo {author}
  {\bibfnamefont {N.-T.}\ \bibnamefont {Nguyen}}, \bibinfo {author}
  {\bibfnamefont {D.}~\bibnamefont {{Di Carlo}}}, \ and\ \bibinfo {author}
  {\bibfnamefont {W.}~\bibnamefont {Li}},\ }\bibfield  {title} {\enquote
  {\bibinfo {title} {Real-time control of inertial focusing in microfluidics
  using dielectrophoresis (dep)},}\ }\href@noop {} {\bibfield  {journal}
  {\bibinfo  {journal} {RSC Adv.}\ }\textbf {\bibinfo {volume} {4}},\ \bibinfo
  {pages} {62076--62085} (\bibinfo {year} {2014})}\BibitemShut {NoStop}%
\bibitem [{\citenamefont {Dutz}\ \emph {et~al.}(2017)\citenamefont {Dutz},
  \citenamefont {Hayden},\ and\ \citenamefont
  {H{\"a}feli}}]{dutz2017fractionation}%
  \BibitemOpen
  \bibfield  {author} {\bibinfo {author} {\bibfnamefont {S.}~\bibnamefont
  {Dutz}}, \bibinfo {author} {\bibfnamefont {M.E.}\ \bibnamefont {Hayden}}, \
  and\ \bibinfo {author} {\bibfnamefont {U.O.}\ \bibnamefont {H{\"a}feli}},\
  }\bibfield  {title} {\enquote {\bibinfo {title} {Fractionation of magnetic
  microspheres in a microfluidic spiral: Interplay between magnetic and
  hydrodynamic forces},}\ }\href@noop {} {\bibfield  {journal} {\bibinfo
  {journal} {PLOS ONE}\ }\textbf {\bibinfo {volume} {12}},\ \bibinfo {pages}
  {e0169919} (\bibinfo {year} {2017})}\BibitemShut {NoStop}%
\bibitem [{\citenamefont {Altena}\ and\ \citenamefont
  {Belfort}(1984)}]{altena1984}%
  \BibitemOpen
  \bibfield  {author} {\bibinfo {author} {\bibfnamefont {F.~W.}\ \bibnamefont
  {Altena}}\ and\ \bibinfo {author} {\bibfnamefont {G.}~\bibnamefont
  {Belfort}},\ }\bibfield  {title} {\enquote {\bibinfo {title} {Lateral
  migration of spherical particles in porous flow channels: application to
  membrane filtration},}\ }\href@noop {} {\bibfield  {journal} {\bibinfo
  {journal} {Chem. Eng. Sci.}\ }\textbf {\bibinfo {volume} {39}},\ \bibinfo
  {pages} {343--355} (\bibinfo {year} {1984})}\BibitemShut {NoStop}%
\bibitem [{\citenamefont {Lebedeva}\ and\ \citenamefont
  {Asmolov}(2011)}]{asmolov2011}%
  \BibitemOpen
  \bibfield  {author} {\bibinfo {author} {\bibfnamefont {N.~A.}\ \bibnamefont
  {Lebedeva}}\ and\ \bibinfo {author} {\bibfnamefont {E.~S.}\ \bibnamefont
  {Asmolov}},\ }\bibfield  {title} {\enquote {\bibinfo {title} {Migration of
  settling particles in a horizontal viscous flow through a vertical slot with
  porous walls},}\ }\href@noop {} {\bibfield  {journal} {\bibinfo  {journal}
  {Int. J. Multiph. Flow}\ }\textbf {\bibinfo {volume} {27}},\ \bibinfo {pages}
  {453--461} (\bibinfo {year} {2011})}\BibitemShut {NoStop}%
\bibitem [{\citenamefont {Garcia}\ and\ \citenamefont
  {Pennathur}(2017)}]{garcia2017}%
  \BibitemOpen
  \bibfield  {author} {\bibinfo {author} {\bibfnamefont {M.}~\bibnamefont
  {Garcia}}\ and\ \bibinfo {author} {\bibfnamefont {S.}~\bibnamefont
  {Pennathur}},\ }\bibfield  {title} {\enquote {\bibinfo {title} {Inertial
  particle dynamics in the presence of a secondary flow},}\ }\href@noop {}
  {\bibfield  {journal} {\bibinfo  {journal} {Phys. Rev. Fluids}\ }\textbf
  {\bibinfo {volume} {2}},\ \bibinfo {pages} {042201} (\bibinfo {year}
  {2017})}\BibitemShut {NoStop}%
\bibitem [{\citenamefont {Quere}(2008)}]{quere.d:2008}%
  \BibitemOpen
  \bibfield  {author} {\bibinfo {author} {\bibfnamefont {D.}~\bibnamefont
  {Quere}},\ }\bibfield  {title} {\enquote {\bibinfo {title} {Wetting and
  roughness},}\ }\href@noop {} {\bibfield  {journal} {\bibinfo  {journal}
  {Annu. Rev. Mater. Res.}\ }\textbf {\bibinfo {volume} {38}},\ \bibinfo
  {pages} {71--99} (\bibinfo {year} {2008})}\BibitemShut {NoStop}%
\bibitem [{\citenamefont {Ybert}\ \emph {et~al.}(2007)\citenamefont {Ybert},
  \citenamefont {Barentin}, \citenamefont {Cottin-Bizonne}, \citenamefont
  {Joseph},\ and\ \citenamefont {Bocquet}}]{ybert.c:2007}%
  \BibitemOpen
  \bibfield  {author} {\bibinfo {author} {\bibfnamefont {C.}~\bibnamefont
  {Ybert}}, \bibinfo {author} {\bibfnamefont {C.}~\bibnamefont {Barentin}},
  \bibinfo {author} {\bibfnamefont {C.}~\bibnamefont {Cottin-Bizonne}},
  \bibinfo {author} {\bibfnamefont {P.}~\bibnamefont {Joseph}}, \ and\ \bibinfo
  {author} {\bibfnamefont {L.}~\bibnamefont {Bocquet}},\ }\bibfield  {title}
  {\enquote {\bibinfo {title} {Achieving large slip with superhydrophobic
  surfaces: Scaling laws for generic geometries},}\ }\href@noop {} {\bibfield
  {journal} {\bibinfo  {journal} {Phys. Fluids}\ }\textbf {\bibinfo {volume}
  {19}},\ \bibinfo {pages} {123601} (\bibinfo {year} {2007})}\BibitemShut
  {NoStop}%
\bibitem [{\citenamefont {Vinogradova}\ and\ \citenamefont
  {Belyaev}(2011)}]{vinogradova.oi:2011}%
  \BibitemOpen
  \bibfield  {author} {\bibinfo {author} {\bibfnamefont {O.~I.}\ \bibnamefont
  {Vinogradova}}\ and\ \bibinfo {author} {\bibfnamefont {A.~V.}\ \bibnamefont
  {Belyaev}},\ }\bibfield  {title} {\enquote {\bibinfo {title} {Wetting,
  roughness and flow boundary conditions},}\ }\href@noop {} {\bibfield
  {journal} {\bibinfo  {journal} {J. Phys.: Condens. Matter}\ }\textbf
  {\bibinfo {volume} {23}},\ \bibinfo {pages} {184104} (\bibinfo {year}
  {2011})}\BibitemShut {NoStop}%
\bibitem [{\citenamefont {Rothstein}(2010)}]{rothstein.jp:2010}%
  \BibitemOpen
  \bibfield  {author} {\bibinfo {author} {\bibfnamefont {J.~P.}\ \bibnamefont
  {Rothstein}},\ }\bibfield  {title} {\enquote {\bibinfo {title} {Slip on
  superhydrophobic surfaces},}\ }\href@noop {} {\bibfield  {journal} {\bibinfo
  {journal} {Annu. Rev. Fluid Mech.}\ }\textbf {\bibinfo {volume} {42}},\
  \bibinfo {pages} {89--109} (\bibinfo {year} {2010})}\BibitemShut {NoStop}%
\bibitem [{\citenamefont {Bazant}\ and\ \citenamefont
  {Vinogradova}(2008)}]{bazant2008}%
  \BibitemOpen
  \bibfield  {author} {\bibinfo {author} {\bibfnamefont {M.~Z.}\ \bibnamefont
  {Bazant}}\ and\ \bibinfo {author} {\bibfnamefont {O.~I.}\ \bibnamefont
  {Vinogradova}},\ }\bibfield  {title} {\enquote {\bibinfo {title} {Tensorial
  hydrodynamic slip},}\ }\href@noop {} {\bibfield  {journal} {\bibinfo
  {journal} {J. Fluid Mech.}\ }\textbf {\bibinfo {volume} {613}},\ \bibinfo
  {pages} {125--134} (\bibinfo {year} {2008})}\BibitemShut {NoStop}%
\bibitem [{\citenamefont {Feuillebois}\ \emph {et~al.}(2009)\citenamefont
  {Feuillebois}, \citenamefont {Bazant},\ and\ \citenamefont
  {Vinogradova}}]{feuillebois.f:2009}%
  \BibitemOpen
  \bibfield  {author} {\bibinfo {author} {\bibfnamefont {F.}~\bibnamefont
  {Feuillebois}}, \bibinfo {author} {\bibfnamefont {M.~Z.}\ \bibnamefont
  {Bazant}}, \ and\ \bibinfo {author} {\bibfnamefont {O.~I.}\ \bibnamefont
  {Vinogradova}},\ }\bibfield  {title} {\enquote {\bibinfo {title} {Effective
  slip over superhydrophobic surfaces in thin channels},}\ }\href@noop {}
  {\bibfield  {journal} {\bibinfo  {journal} {Phys. Rev. Lett.}\ }\textbf
  {\bibinfo {volume} {102}},\ \bibinfo {pages} {026001} (\bibinfo {year}
  {2009})}\BibitemShut {NoStop}%
\bibitem [{\citenamefont {Schmieschek}\ \emph {et~al.}(2012)\citenamefont
  {Schmieschek}, \citenamefont {Belyaev}, \citenamefont {Harting},\ and\
  \citenamefont {Vinogradova}}]{schmieschek.s:2012}%
  \BibitemOpen
  \bibfield  {author} {\bibinfo {author} {\bibfnamefont {S.}~\bibnamefont
  {Schmieschek}}, \bibinfo {author} {\bibfnamefont {A.~V.}\ \bibnamefont
  {Belyaev}}, \bibinfo {author} {\bibfnamefont {J.}~\bibnamefont {Harting}}, \
  and\ \bibinfo {author} {\bibfnamefont {O.~I.}\ \bibnamefont {Vinogradova}},\
  }\bibfield  {title} {\enquote {\bibinfo {title} {Tensorial slip of
  super-hydrophobic channels},}\ }\href@noop {} {\bibfield  {journal} {\bibinfo
   {journal} {Phys. Rev. E}\ }\textbf {\bibinfo {volume} {85}},\ \bibinfo
  {pages} {016324} (\bibinfo {year} {2012})}\BibitemShut {NoStop}%
\bibitem [{\citenamefont {Feuillebois}\ \emph {et~al.}(2010)\citenamefont
  {Feuillebois}, \citenamefont {Bazant},\ and\ \citenamefont
  {Vinogradova}}]{feuillebois.f:2010b}%
  \BibitemOpen
  \bibfield  {author} {\bibinfo {author} {\bibfnamefont {F.}~\bibnamefont
  {Feuillebois}}, \bibinfo {author} {\bibfnamefont {M.~Z.}\ \bibnamefont
  {Bazant}}, \ and\ \bibinfo {author} {\bibfnamefont {O.~I.}\ \bibnamefont
  {Vinogradova}},\ }\bibfield  {title} {\enquote {\bibinfo {title} {Transverse
  flow in thin superhydrophobic channels},}\ }\href@noop {} {\bibfield
  {journal} {\bibinfo  {journal} {Phys. Rev. E}\ }\textbf {\bibinfo {volume}
  {82}},\ \bibinfo {pages} {055301(R)} (\bibinfo {year} {2010})}\BibitemShut
  {NoStop}%
\bibitem [{\citenamefont {Pimponi}\ \emph {et~al.}(2014)\citenamefont
  {Pimponi}, \citenamefont {Chinappi}, \citenamefont {Gualtieri},\ and\
  \citenamefont {Casciola}}]{pimponi.d:2014}%
  \BibitemOpen
  \bibfield  {author} {\bibinfo {author} {\bibfnamefont {D.}~\bibnamefont
  {Pimponi}}, \bibinfo {author} {\bibfnamefont {M.}~\bibnamefont {Chinappi}},
  \bibinfo {author} {\bibfnamefont {P.}~\bibnamefont {Gualtieri}}, \ and\
  \bibinfo {author} {\bibfnamefont {C.~M.}\ \bibnamefont {Casciola}},\
  }\bibfield  {title} {\enquote {\bibinfo {title} {Mobility tensor of a sphere
  moving on a superhydrophobic wall: application to particle separation},}\
  }\href@noop {} {\bibfield  {journal} {\bibinfo  {journal} {Microfluid
  Nanofluid}\ }\textbf {\bibinfo {volume} {16}},\ \bibinfo {pages} {571--585}
  (\bibinfo {year} {2014})}\BibitemShut {NoStop}%
\bibitem [{\citenamefont {Asmolov}\ \emph {et~al.}(2015)\citenamefont
  {Asmolov}, \citenamefont {Dubov}, \citenamefont {Nizkaya}, \citenamefont
  {Kuehne},\ and\ \citenamefont {Vinogradova}}]{asmolov_LabChip}%
  \BibitemOpen
  \bibfield  {author} {\bibinfo {author} {\bibfnamefont {E.~S.}\ \bibnamefont
  {Asmolov}}, \bibinfo {author} {\bibfnamefont {A.~L.}\ \bibnamefont {Dubov}},
  \bibinfo {author} {\bibfnamefont {T.~V.}\ \bibnamefont {Nizkaya}}, \bibinfo
  {author} {\bibfnamefont {A.~J.~C.}\ \bibnamefont {Kuehne}}, \ and\ \bibinfo
  {author} {\bibfnamefont {O.~I.}\ \bibnamefont {Vinogradova}},\ }\bibfield
  {title} {\enquote {\bibinfo {title} {Principles of transverse flow
  fractionation of microparticles in superhydrophobic channels},}\ }\href@noop
  {} {\bibfield  {journal} {\bibinfo  {journal} {Lab. Chip}\ }\textbf {\bibinfo
  {volume} {15}},\ \bibinfo {pages} {2835--2841} (\bibinfo {year}
  {2015})}\BibitemShut {NoStop}%
\bibitem [{\citenamefont {Sbragaglia}\ and\ \citenamefont
  {Prosperetti}(2007)}]{sbragaglia.m:2007}%
  \BibitemOpen
  \bibfield  {author} {\bibinfo {author} {\bibfnamefont {M.}~\bibnamefont
  {Sbragaglia}}\ and\ \bibinfo {author} {\bibfnamefont {A.}~\bibnamefont
  {Prosperetti}},\ }\bibfield  {title} {\enquote {\bibinfo {title} {A note on
  the effective slip properties for microchannel flows with ultrahydrophobic
  surfaces},}\ }\href@noop {} {\bibfield  {journal} {\bibinfo  {journal} {Phys.
  Fluids}\ }\textbf {\bibinfo {volume} {19}},\ \bibinfo {pages} {043603}
  (\bibinfo {year} {2007})}\BibitemShut {NoStop}%
\bibitem [{\citenamefont {Teo}\ and\ \citenamefont {Khoo}(2009)}]{Teo&Khoo:09}%
  \BibitemOpen
  \bibfield  {author} {\bibinfo {author} {\bibfnamefont {C.}~\bibnamefont
  {Teo}}\ and\ \bibinfo {author} {\bibfnamefont {B.}~\bibnamefont {Khoo}},\
  }\bibfield  {title} {\enquote {\bibinfo {title} {Analysis of {Stokes} flow in
  microchannels with superhydrophobic surfaces containing a periodic array of
  micro-grooves},}\ }\href@noop {} {\bibfield  {journal} {\bibinfo  {journal}
  {Microfluid Nanofluid}\ }\textbf {\bibinfo {volume} {7}},\ \bibinfo {pages}
  {353} (\bibinfo {year} {2009})}\BibitemShut {NoStop}%
\bibitem [{\citenamefont {Asmolov}\ \emph
  {et~al.}(2018{\natexlab{a}})\citenamefont {Asmolov}, \citenamefont
  {Nizkaya},\ and\ \citenamefont {Vinogradova}}]{nizkaya.tv:2018}%
  \BibitemOpen
  \bibfield  {author} {\bibinfo {author} {\bibfnamefont {E.~S.}\ \bibnamefont
  {Asmolov}}, \bibinfo {author} {\bibfnamefont {T.~V.}\ \bibnamefont
  {Nizkaya}}, \ and\ \bibinfo {author} {\bibfnamefont {O.~I.}\ \bibnamefont
  {Vinogradova}},\ }\bibfield  {title} {\enquote {\bibinfo {title} {Enhanced
  slip properties of lubricant-infused grooves},}\ }\href@noop {} {\bibfield
  {journal} {\bibinfo  {journal} {Phys. Rev. E}\ }\textbf {\bibinfo {volume}
  {98}},\ \bibinfo {pages} {033103} (\bibinfo {year}
  {2018}{\natexlab{a}})}\BibitemShut {NoStop}%
\bibitem [{\citenamefont {Asmolov}\ \emph
  {et~al.}(2018{\natexlab{b}})\citenamefont {Asmolov}, \citenamefont {Dubov},
  \citenamefont {Nizkaya}, \citenamefont {Harting},\ and\ \citenamefont
  {Vinogradova}}]{asmolov2018}%
  \BibitemOpen
  \bibfield  {author} {\bibinfo {author} {\bibfnamefont {E.~S.}\ \bibnamefont
  {Asmolov}}, \bibinfo {author} {\bibfnamefont {A.~L.}\ \bibnamefont {Dubov}},
  \bibinfo {author} {\bibfnamefont {T.~V.}\ \bibnamefont {Nizkaya}}, \bibinfo
  {author} {\bibfnamefont {J.}~\bibnamefont {Harting}}, \ and\ \bibinfo
  {author} {\bibfnamefont {O.~I.}\ \bibnamefont {Vinogradova}},\ }\bibfield
  {title} {\enquote {\bibinfo {title} {Inertial focusing of finite-size
  particles in microchannels},}\ }\href@noop {} {\bibfield  {journal} {\bibinfo
   {journal} {J. Fluid Mech.}\ }\textbf {\bibinfo {volume} {840}},\ \bibinfo
  {pages} {613--630} (\bibinfo {year} {2018}{\natexlab{b}})}\BibitemShut
  {NoStop}%
\bibitem [{\citenamefont {Davis}\ \emph {et~al.}(1994)\citenamefont {Davis},
  \citenamefont {Kezirian},\ and\ \citenamefont {Brenner}}]{davis1994}%
  \BibitemOpen
  \bibfield  {author} {\bibinfo {author} {\bibfnamefont {A.~M.~J.}\
  \bibnamefont {Davis}}, \bibinfo {author} {\bibfnamefont {M.~T.}\ \bibnamefont
  {Kezirian}}, \ and\ \bibinfo {author} {\bibfnamefont {H.}~\bibnamefont
  {Brenner}},\ }\bibfield  {title} {\enquote {\bibinfo {title} {On the
  {Stokes-Einstein} model of surface diffusion along solid surfaces: {Slip}
  boundary conditions},}\ }\href@noop {} {\bibfield  {journal} {\bibinfo
  {journal} {J. Colloid Interface Sci.}\ }\textbf {\bibinfo {volume} {165}},\
  \bibinfo {pages} {129--140} (\bibinfo {year} {1994})}\BibitemShut {NoStop}%
\bibitem [{\citenamefont {Loussaief}\ \emph {et~al.}(2015)\citenamefont
  {Loussaief}, \citenamefont {Pasol},\ and\ \citenamefont
  {Feuillebois}}]{loussaief15}%
  \BibitemOpen
  \bibfield  {author} {\bibinfo {author} {\bibfnamefont {H.}~\bibnamefont
  {Loussaief}}, \bibinfo {author} {\bibfnamefont {L.}~\bibnamefont {Pasol}}, \
  and\ \bibinfo {author} {\bibfnamefont {F.}~\bibnamefont {Feuillebois}},\
  }\bibfield  {title} {\enquote {\bibinfo {title} {Motion of a spherical
  particle in a viscous fluid along a slip wall},}\ }\href@noop {} {\bibfield
  {journal} {\bibinfo  {journal} {Q. J. Mech. Appl. Math.}\ }\textbf {\bibinfo
  {volume} {68}},\ \bibinfo {pages} {115--144} (\bibinfo {year}
  {2015})}\BibitemShut {NoStop}%
\bibitem [{\citenamefont {Ghalia}\ \emph {et~al.}(2016)\citenamefont {Ghalia},
  \citenamefont {Feuillebois},\ and\ \citenamefont {Sellier}}]{ghalia2016}%
  \BibitemOpen
  \bibfield  {author} {\bibinfo {author} {\bibfnamefont {N.}~\bibnamefont
  {Ghalia}}, \bibinfo {author} {\bibfnamefont {F.}~\bibnamefont {Feuillebois}},
  \ and\ \bibinfo {author} {\bibfnamefont {A.}~\bibnamefont {Sellier}},\
  }\bibfield  {title} {\enquote {\bibinfo {title} {A sphere in a second degree
  polynomial creeping flow parallel to a plane, impermeable and slipping
  wall},}\ }\href@noop {} {\bibfield  {journal} {\bibinfo  {journal} {Q. J.
  Mech. Appl. Math.}\ }\textbf {\bibinfo {volume} {69}},\ \bibinfo {pages}
  {353--390} (\bibinfo {year} {2016})}\BibitemShut {NoStop}%
\bibitem [{\citenamefont {Asmolov}(1999)}]{Asmolov99}%
  \BibitemOpen
  \bibfield  {author} {\bibinfo {author} {\bibfnamefont {E.~S.}\ \bibnamefont
  {Asmolov}},\ }\bibfield  {title} {\enquote {\bibinfo {title} {The inertial
  lift on a spherical particle in a plane {Poiseuille} flow at large channel
  {Reynolds} number},}\ }\href@noop {} {\bibfield  {journal} {\bibinfo
  {journal} {J. Fluid Mech.}\ }\textbf {\bibinfo {volume} {381}},\ \bibinfo
  {pages} {63--87} (\bibinfo {year} {1999})}\BibitemShut {NoStop}%
\bibitem [{\citenamefont {Benzi}\ \emph {et~al.}(1992)\citenamefont {Benzi},
  \citenamefont {Succi},\ and\ \citenamefont
  {Vergassola}}]{benzi_lattice_1992}%
  \BibitemOpen
  \bibfield  {author} {\bibinfo {author} {\bibfnamefont {R.}~\bibnamefont
  {Benzi}}, \bibinfo {author} {\bibfnamefont {S.}~\bibnamefont {Succi}}, \ and\
  \bibinfo {author} {\bibfnamefont {M.}~\bibnamefont {Vergassola}},\ }\bibfield
   {title} {\enquote {\bibinfo {title} {The lattice {B}oltzmann equation:
  theory and applications},}\ }\href@noop {} {\bibfield  {journal} {\bibinfo
  {journal} {Physics Reports}\ }\textbf {\bibinfo {volume} {222}},\ \bibinfo
  {pages} {145} (\bibinfo {year} {1992})}\BibitemShut {NoStop}%
\bibitem [{\citenamefont {Kunert}\ \emph {et~al.}(2010)\citenamefont {Kunert},
  \citenamefont {Harting},\ and\ \citenamefont
  {Vinogradova}}]{kunert2010random}%
  \BibitemOpen
  \bibfield  {author} {\bibinfo {author} {\bibfnamefont {C.}~\bibnamefont
  {Kunert}}, \bibinfo {author} {\bibfnamefont {J.}~\bibnamefont {Harting}}, \
  and\ \bibinfo {author} {\bibfnamefont {O.~I.}\ \bibnamefont {Vinogradova}},\
  }\bibfield  {title} {\enquote {\bibinfo {title} {Random-roughness
  hydrodynamic boundary conditions},}\ }\href@noop {} {\bibfield  {journal}
  {\bibinfo  {journal} {Phys. Rev. Lett.}\ }\textbf {\bibinfo {volume} {105}},\
  \bibinfo {pages} {016001} (\bibinfo {year} {2010})}\BibitemShut {NoStop}%
\bibitem [{\citenamefont {Ladd}\ and\ \citenamefont
  {Verberg}(2001)}]{LaddVerberg2001}%
  \BibitemOpen
  \bibfield  {author} {\bibinfo {author} {\bibfnamefont {A.~J.~C.}\
  \bibnamefont {Ladd}}\ and\ \bibinfo {author} {\bibfnamefont {R.}~\bibnamefont
  {Verberg}},\ }\bibfield  {title} {\enquote {\bibinfo {title}
  {Lattice-{Boltzmann} simulations of particle-fluid suspensions},}\
  }\href@noop {} {\bibfield  {journal} {\bibinfo  {journal} {J. Stat. Phys.}\
  }\textbf {\bibinfo {volume} {104}},\ \bibinfo {pages} {1191} (\bibinfo {year}
  {2001})}\BibitemShut {NoStop}%
\bibitem [{\citenamefont {Janoschek}\ \emph {et~al.}(2010)\citenamefont
  {Janoschek}, \citenamefont {Toschi},\ and\ \citenamefont
  {Harting}}]{bib:jens-janoschek-toschi-2010b}%
  \BibitemOpen
  \bibfield  {author} {\bibinfo {author} {\bibfnamefont {F.}~\bibnamefont
  {Janoschek}}, \bibinfo {author} {\bibfnamefont {F.}~\bibnamefont {Toschi}}, \
  and\ \bibinfo {author} {\bibfnamefont {J.}~\bibnamefont {Harting}},\
  }\bibfield  {title} {\enquote {\bibinfo {title} {Simplified particulate model
  for coarse-grained hemodynamics simulations},}\ }\href@noop {} {\bibfield
  {journal} {\bibinfo  {journal} {Phys. Rev. E}\ }\textbf {\bibinfo {volume}
  {82}},\ \bibinfo {pages} {056710} (\bibinfo {year} {2010})}\BibitemShut
  {NoStop}%
\bibitem [{\citenamefont {Janoschek}\ \emph {et~al.}(2014)\citenamefont
  {Janoschek}, \citenamefont {Harting},\ and\ \citenamefont {Toschi}}]{JHT14a}%
  \BibitemOpen
  \bibfield  {author} {\bibinfo {author} {\bibfnamefont {F.}~\bibnamefont
  {Janoschek}}, \bibinfo {author} {\bibfnamefont {J.}~\bibnamefont {Harting}},
  \ and\ \bibinfo {author} {\bibfnamefont {F.}~\bibnamefont {Toschi}},\
  }\bibfield  {title} {\enquote {\bibinfo {title} {Towards a continuum model
  for particle-induced velocity fluctuations in suspension flow through a
  stenosed geometry},}\ }\href@noop {} {\bibfield  {journal} {\bibinfo
  {journal} {Int. J. Modern Physics C}\ }\textbf {\bibinfo {volume} {25}},\
  \bibinfo {pages} {1441013} (\bibinfo {year} {2014})}\BibitemShut {NoStop}%
\bibitem [{\citenamefont {Dubov}\ \emph {et~al.}(2014)\citenamefont {Dubov},
  \citenamefont {Schmieschek}, \citenamefont {Asmolov}, \citenamefont
  {Harting},\ and\ \citenamefont {Vinogradova}}]{DSAHV14}%
  \BibitemOpen
  \bibfield  {author} {\bibinfo {author} {\bibfnamefont {A.~L.}\ \bibnamefont
  {Dubov}}, \bibinfo {author} {\bibfnamefont {S.}~\bibnamefont {Schmieschek}},
  \bibinfo {author} {\bibfnamefont {E.~S.}\ \bibnamefont {Asmolov}}, \bibinfo
  {author} {\bibfnamefont {J.}~\bibnamefont {Harting}}, \ and\ \bibinfo
  {author} {\bibfnamefont {O.I.}\ \bibnamefont {Vinogradova}},\ }\bibfield
  {title} {\enquote {\bibinfo {title} {Lattice-{B}oltzmann simulations of the
  drag force on a sphere approaching a superhydrophobic striped plane},}\
  }\href@noop {} {\bibfield  {journal} {\bibinfo  {journal} {J. Chem. Phys.}\
  }\textbf {\bibinfo {volume} {140}},\ \bibinfo {pages} {034707} (\bibinfo
  {year} {2014})}\BibitemShut {NoStop}%
\bibitem [{\citenamefont {Harting}\ \emph {et~al.}(2014)\citenamefont
  {Harting}, \citenamefont {Frijters}, \citenamefont {Ramaioli}, \citenamefont
  {Robinson}, \citenamefont {Wolf},\ and\ \citenamefont {Luding}}]{HFRRWL14}%
  \BibitemOpen
  \bibfield  {author} {\bibinfo {author} {\bibfnamefont {J.}~\bibnamefont
  {Harting}}, \bibinfo {author} {\bibfnamefont {S.}~\bibnamefont {Frijters}},
  \bibinfo {author} {\bibfnamefont {M.}~\bibnamefont {Ramaioli}}, \bibinfo
  {author} {\bibfnamefont {M.}~\bibnamefont {Robinson}}, \bibinfo {author}
  {\bibfnamefont {D.~E.}\ \bibnamefont {Wolf}}, \ and\ \bibinfo {author}
  {\bibfnamefont {S.}~\bibnamefont {Luding}},\ }\bibfield  {title} {\enquote
  {\bibinfo {title} {Recent advances in the simulation of particle-laden
  flows},}\ }\href@noop {} {\bibfield  {journal} {\bibinfo  {journal} {Eur.
  Phys. J. Spec. Topics}\ }\textbf {\bibinfo {volume} {223}},\ \bibinfo {pages}
  {2253--2267} (\bibinfo {year} {2014})}\BibitemShut {NoStop}%
\bibitem [{\citenamefont {Ahmed}\ and\ \citenamefont
  {Hecht}(2009)}]{AhmedHecht2009}%
  \BibitemOpen
  \bibfield  {author} {\bibinfo {author} {\bibfnamefont {N.~K.}\ \bibnamefont
  {Ahmed}}\ and\ \bibinfo {author} {\bibfnamefont {M.}~\bibnamefont {Hecht}},\
  }\bibfield  {title} {\enquote {\bibinfo {title} {A boundary condition with
  adjustable slip length for lattice boltzmann simulations},}\ }\href
  {http://stacks.iop.org/1742-5468/2009/i=09/a=P09017} {\bibfield  {journal}
  {\bibinfo  {journal} {J. Stat. Mech. Theory Exp.}\ }\textbf {\bibinfo
  {volume} {2009}},\ \bibinfo {pages} {09017} (\bibinfo {year}
  {2009})}\BibitemShut {NoStop}%
\bibitem [{\citenamefont {Hecht}\ and\ \citenamefont
  {Harting}(2010)}]{HechtHarting2010}%
  \BibitemOpen
  \bibfield  {author} {\bibinfo {author} {\bibfnamefont {M.}~\bibnamefont
  {Hecht}}\ and\ \bibinfo {author} {\bibfnamefont {J.}~\bibnamefont
  {Harting}},\ }\bibfield  {title} {\enquote {\bibinfo {title} {Implementation
  of on-site velocity boundary conditions for d3q19 lattice boltzmann
  simulations},}\ }\href {http://stacks.iop.org/1742-5468/2010/i=01/a=P01018}
  {\bibfield  {journal} {\bibinfo  {journal} {J. Stat. Mech. Theory Exp.}\
  }\textbf {\bibinfo {volume} {2010}},\ \bibinfo {pages} {P01018} (\bibinfo
  {year} {2010})}\BibitemShut {NoStop}%
\bibitem [{\citenamefont {Nizkaya}\ \emph {et~al.}(2013)\citenamefont
  {Nizkaya}, \citenamefont {Asmolov},\ and\ \citenamefont
  {Vinogradova}}]{nizkaya2013}%
  \BibitemOpen
  \bibfield  {author} {\bibinfo {author} {\bibfnamefont {T.~V.}\ \bibnamefont
  {Nizkaya}}, \bibinfo {author} {\bibfnamefont {E.~S.}\ \bibnamefont
  {Asmolov}}, \ and\ \bibinfo {author} {\bibfnamefont {O.~I.}\ \bibnamefont
  {Vinogradova}},\ }\bibfield  {title} {\enquote {\bibinfo {title} {Flow in
  channels with superhydrophobic trapezoidal textures},}\ }\href {\doibase
  10.1039/C3SM51850G} {\bibfield  {journal} {\bibinfo  {journal} {Soft Matter}\
  }\textbf {\bibinfo {volume} {9}},\ \bibinfo {pages} {11671--11679} (\bibinfo
  {year} {2013})}\BibitemShut {NoStop}%
\bibitem [{\citenamefont {Nizkaya}\ \emph {et~al.}(2015)\citenamefont
  {Nizkaya}, \citenamefont {Asmolov}, \citenamefont {Zhou}, \citenamefont
  {Schmid},\ and\ \citenamefont {Vinogradova}}]{nizkaya2015}%
  \BibitemOpen
  \bibfield  {author} {\bibinfo {author} {\bibfnamefont {T.~V.}\ \bibnamefont
  {Nizkaya}}, \bibinfo {author} {\bibfnamefont {E.~S.}\ \bibnamefont
  {Asmolov}}, \bibinfo {author} {\bibfnamefont {J.}~\bibnamefont {Zhou}},
  \bibinfo {author} {\bibfnamefont {F.}~\bibnamefont {Schmid}}, \ and\ \bibinfo
  {author} {\bibfnamefont {O.~I.}\ \bibnamefont {Vinogradova}},\ }\bibfield
  {title} {\enquote {\bibinfo {title} {Flows and mixing in channels with
  misaligned superhydrophobic walls},}\ }\href@noop {} {\bibfield  {journal}
  {\bibinfo  {journal} {Phys. Rev. E}\ }\textbf {\bibinfo {volume} {91}},\
  \bibinfo {pages} {033020} (\bibinfo {year} {2015})}\BibitemShut {NoStop}%
\end{thebibliography}%

\end{document}